\documentclass[12pt,preprint]{aastex}

\def\kms{\rm \, km \, s^{-1}}
\def\msun{M_{\sun}}

\shorttitle{The 25 Orionis group in Orion OB1a}
\shortauthors{Brice\~no et al.}

\begin{document}


\title{25 Orionis: A kinematically distinct 10 Myr old group in Orion OB1a\altaffilmark{1}}


\author{C\'esar Brice\~no\altaffilmark{2,5}, 
Lee Hartmann\altaffilmark{3}, Jes\'us Hern\'andez\altaffilmark{2,3}, 
Nuria Calvet\altaffilmark{3},
A. Katherina Vivas\altaffilmark{2}, Gabor Furesz\altaffilmark{4}, 
Andrew Szentgyorgyi\altaffilmark{4}}

\altaffiltext{1}{Based on observations obtained at the
Llano del Hato National Astronomical Observatory of Venezuela,
operated by CIDA for the Ministerio de Ciencia y Tecnolog{\'\i}a;
the MMT Observatory, a joint facility of the Smithsonian
Institution and the University of Arizona; and the Fred Lawrence Whipple
Observatory of the Smithsonian Institution, USA}
\altaffiltext{2}{Centro de Investigaciones de Astronom{\'\i}a (CIDA), 
Apartado Postal 264, M\'erida, 5101-A, Venezuela. E-mails: briceno@cida.ve, akvivas@cida.ve}
\altaffiltext{3}{Department of Astronomy, University of Michigan,
500 Church St. 830 Dennison, Ann Arbor, MI 48109, USA. E-mails: lhartm@umich.edu, ncalvet@umich.edu, hernandj@umich.edu}
\altaffiltext{4}{Harvard-Smithsonian Center for Astrophysics, 60 Garden St., Cambridge, MA 02138, USA. E-mails: gfuresz@cfa.harvard.edu, saint@cfa.harvard.edu}
\altaffiltext{5}{Associate Researcher, Smithsonian Astrophysical Observatory}

\email{briceno@cida.ve}



\begin{abstract}
We report here on the photometric and kinematic properties of 
a well defined group of nearly 200 low-mass pre-main sequence stars, 
concentrated within $\sim 1\arcdeg$ of the early-B star 25 Ori,
in the Orion OB1a sub-association. 
We refer to this stellar aggregate as the 25 Orionis group.
The group also harbors the Herbig Ae/Be star V346 Ori
and a dozen other early type stars with photometry, parallaxes, and some
with IR excess emission, consistent with group membership. 
The number of high and low-mass stars is in agreement
with expectations from a standard Initial Mass Function.
The velocity distribution for the young stars in 25 Ori shows a narrow
peak centered at $19.7 \kms$, very close to the velocity of the star 25 Ori.
This velocity peak is offset $\sim -10 \kms$ from the velocity characterizing
the younger stars of the Ori OB1b sub-association, and  $-4 \kms$ from the
velocity of more widely spread young stars of the
Ori OB1a population near the $\sigma$ Ori cluster; this result
provides new and compelling 
evidence that the 25 Ori group is a distinct kinematic entity,
and that considerable space and velocity structure is present
in the Ori OB1a sub-association.
The low-mass members follow a well defined band in the color-magnitude
diagram, consistent with an isochronal age of $\sim 7-10$ Myr,
depending on the assumed evolutionary model.
The $\sim 2\times $ drop in the overall Li~I equivalent widths and 
accretion fraction between the younger Ori OB1b and the 25 Ori group,
is consistent with the later being significantly older,
independent of the absolute age calibration.\\
The highest density of members is located near the
star 25 Ori, but the actual extent of the cluster cannot be
well constrained with our present data. 
In a simple-minded kinematic evolution scenario, the 25 Ori group 
may represent the evolved counterpart of a younger
aggregate like the $\sigma$ Ori cluster. \\
The 25 Ori stellar aggregate is the most populous $\sim 10$ Myr sample yet known
within 500 pc, setting it as an excellent laboratory to study 
the evolution of solar-like stars and protoplanetary disks.
\end{abstract}





\section{Introduction}

Samples of $\sim 10$ Myr old, low-mass 
stars are important for understanding protoplanetary disk
and pre-main sequence stellar evolution. 
Many estimates \citep[e.g.][]{briceno01,calvet05,haisch01}
indicate that by $\la 10$ Myr the majority of inner disks have dissipated
and planet formation has been mostly completed \citep{pollack96}.
However, finding large samples of stars at this age
has been a significant observational challenge. 

The traditional means of identifying low-mass
pre-main sequence (PMS) stars (or T Tauri stars - TTS, as they
are usually known) has been to look toward molecular clouds. 
If cloud complexes had long lifetimes, of $\sim 10-100$ Myr,
\citep[e.g.][]{pollack96},
and star formation had proceeded in a more or less continuous fashion, 
one would expect to find roughly 10 times as many 10 Myr-old stars as 1 Myr old
objects toward dark clouds with crossing times of order 10 Myr. 
The lack of such older PMS stars in regions like the Taurus clouds
was called by \cite{herbig78} the Post-T Tauri problem.
Recently, accumulating observational and theoretical evidence suggest that
molecular clouds lifetimes are much shorter than previously thought
\citep[e.g.][]{ballesteros06,briceno06}; 
by ages $\la 10$ Myr the parent molecular clouds 
have dissipated and no longer serve as markers of the young population 
\citep[c.f.][]{briceno05}. This is consistent with the absence of
$\rm ^{12}CO$ emission in regions like the Orion OB1a sub-association,
in which the estimated ages of the massive OBA stars are of order 10 Myr
\citep{brown94,dezeeuw99}.
One approach to searching for somewhat older PMS stars over wide areas,
extending beyond the confines of molecular clouds,
has been to  select X-ray active G and K type stars in
large scale X-ray surveys like the ROSAT All-Sky survey (RASS). However,
in regions like Orion the RASS only reached down to $\sim 1 \msun$ stars \citep{alcala96},
missing the more numerous late K and M type stars; in addition,
it suffered from significant contamination by foreground X-ray active
zero age main sequence stars \citep{briceno97}.

Nevertheless, a few $\sim 10$ Myr old stellar groups have been discovered in the solar
neighborhood in recent years.
The nearby TW Hya association is one of the most important such groups,
but the coevality or even membership is somewhat uncertain for some
objects \citep{mamajek05}. Another important 
nearby group of similar age is the $\eta$ Cha
cluster \citep{mamajek99} but, like TW Hya, it contains no more than 
$\sim 20$ stars. 
\cite{sicilia05} found a 10 Myr old sample ($\sim 55$ members)
in NGC 7160 in Cepheus, but this is a much more distant cluster ($\sim 800$ pc).

The Orion OB1a sub-association \cite{blaauw64}, 
at an estimated of $\sim 8-10$ Myr \cite{briceno05,briceno06} 
is a promising region for searching for numerous samples of low-mass stars at this
important age range.
In \cite{briceno05} we presented the first sample 
of the low-mass stellar population in this region, 
and identified a concentration of PMS stars around the B1V star 25 Ori.
Here we investigate the kinematics of members of this 
stellar aggregate, which we refer to as the 25 Orionis group.
We show that the stars in this region have a relatively narrow velocity 
dispersion that is kinematically distinct from the stars in the Ori OB1b
sub-association. The location of these stars in the color-magnitude diagram 
indicates a small dispersion in age, and provides indications of systematic
discrepancies in pre-main sequence isochrones.
The stars in the 25 Ori group provide a well defined, populous, sample 
for studies of disk evolution. In section \ref{obs} we describe the
photometric selection of candidate PMS stars, and the spectroscopic
follow-up for membership confirmation, and measurement of radial velocities.
In \S \ref{results} we present our results, describing the overall 25 Ori
stellar population (\S \ref{population}), kinematics (\S \ref{kinematics},
age (\S \ref{ages}), the behavior of Li I as a youth indicator (\S \ref{lithium}),
and the fraction of stars showing evidence for disk accretion (\S \ref{accretion}).

\section{Observations}\label{obs}

\subsection{Selection of T Tauri candidates}\label{selection}

In order to study the star-forming history and characterize the
stellar population
of the Orion OB1 association we are carrying out a large
scale optical variability survey and spectroscopic follow-up.
In Brice\~no et al. (2005) we presented results for a region 
spanning $\rm \sim 68\> deg^2$, between $\alpha_{2000} = 75\arcdeg - 90\arcdeg $
and $\delta_{2000}= -2.13\arcdeg - +2.13\arcdeg$ (in Figure \ref{fig1}
we show a part of that larger area), 
which includes the Ori OB1b sub-association, and large fraction of Ori OB1a.
The multi-band, multi-epoch photometric survey,
has been done with the 
$8000 \times 8000$  QuEST CCD Mosaic Camera \citep{baltay02}
installed on the Stock 1m Schmidt-type telescope at the 
Venezuela Llano del Hato National Astronomical Observatory.
We use variability and optical colors
as a means to select candidate PMS low-mass stars
in color-magnitude diagrams, down to a limiting magnitude of
V$\sim 19.7$. We refer the reader to \cite{briceno05}
for details of the instrument, photometry and
candidate selection criteria.

The photometric sample considered here is composed of
candidate PMS stars located in the same area of the 
\cite{briceno05} study, for which spectroscopic
confirmation is presented here for the first time, plus
stars already confirmed spectroscopically as Orion OB1
members in that work.

\subsection{Spectroscopy}\label{spectroscopy}

\subsubsection{Low resolution spectroscopy: confirming membership}\label{lowres}

Low-resolution spectra provide confirmation of the TTS nature
of a star.
Such objects can be reliably identified through optical spectroscopy
by the presence of: (1) H$\alpha$ emission,
strong in stars actively accreting from
their surrounding disks (the so called Classical T Tauri stars - CTTS,
which exhibit W(H$\alpha$)$ \ge 10${\AA } at M0, with larger values at
later spectral types, see \cite{wba03}), 
and weak (a few \AA) in non-accreting stars, the Weak-lined T Tauri stars - WTTS;
(2) the Li~ I $\lambda 6707${\AA } line
strongly in absorption \citep[e.g.][]{briceno97}.
Spectra in the range $\sim 4000-9000${\AA } also include
several strong molecular features like TiO bands,
characteristic of cooler stars (spectral types later than $\sim K5$), 
which are sensitive T$_{\rm eff}$ indicators.

We obtained low-resolution optical spectra of our candidate PMS stars
with a variety of telescopes and instruments. 
The brighter stars (V$\la 16$) were
observed at the 1.5 m telescope of the
Whipple Observatory equipped with the FAST Spectrograph \citep{fabricant98},
with a spectral coverage of 3400 {\AA } centered
at 5500 {\AA}, and a resolution of 6 {\AA }.
Though spectra for a large fraction of the bright candidates were obtained
in \cite{briceno05}, here we include a number of stars
for which spectra were not available at the time. With these additions
we consider our follow-up spectroscopy in this magnitude range complete.

For the fainter (V$\ga 16$) candidates in our photometric survey
the bulk of the spectroscopic follow-up was done with
the Hectospec multi-fiber spectrograph on the 
6.5 m MMT telescope at Mount Hopkins \citep{fabricant05}. 
Hectospec is a low-resolution optical spectrograph with 
a total of 300 fibers that can be placed within a $1^\circ$ diameter field. We used the 
270 groove $mm^{-1}$ grating, obtaining spectra in the range $\lambda 3700 - 9000 $\AA, with a 
resolution of 6.2 \AA. 
With a S/N ratio $\ga 20$ in our spectra we could detect Li $\lambda 6707$ 
absorption down to $\rm W(Li~ I) \sim 0.2${\AA } to identify the low-mass, young stars.
We measured H$\alpha$ and Li~I equivalent widths in all our low resolution spectra,
using the {\sl splot} routine in IRAF.
From November 2004 through April 2005 we observed 2216 candidates in
eight fields distributed throughout the Orion 1a and 1b sub-associations;
a complete analysis of this entire sample will be presented in
Brice\~no et al. (2007). 
In this contribution we concentrate on newly identified
TTS in a smaller area surrounding the early type star 25 Orionis, in the Orion OB1a
region, and near the belt star $\epsilon$ Ori, located in the Orion OB1b sub-association;
for this sample we present high-resolution spectroscopy (\S \ref{hectochelle}).
The data reduction of both FAST and Hectospec data 
was performed by S. Tokarz through the CfA Telescope 
Data Center, using IRAF tasks and other customized reduction scripts. 
Spectral types were derived using the SPTCLASS code developed at CIDA, 
which includes spectral indices in the Hectospec wavelength range.

\subsubsection{Hectochelle Spectroscopy: investigating kinematics}\label{hectochelle}

In order to study the kinematics of the region, we obtained high resolution 
spectra of a subset of Orion OB1 candidate members,
using the Hectochelle fiber-fed multi-object echelle spectrograph,
mounted on the 6.5m MMT \citep{sze98} at the MMTO, Arizona, USA. 
Hectochelle uses 240 fibers in a $1^\circ$ circular field. 
Each fiber has a diameter of $250\> \mu$m, subtending 1.5 arc sec on the sky.
We observed in the 
spectral order centered near H$\alpha$, from 6465{\AA } to 6650{\AA} (185{\AA } wide),
with a resolution R$\sim 34000$ (equivalent to $\sim 0.19${\AA } FWHM). 
Two observations were obtained on the night of December 2, 2004, 
in two fields containing a total of 243 T Tauri stars
confirmed through FAST and Hectospec spectra, out of which we were
able to obtain high-resolution spectra for 147 objects (60\%).
The first field observed (Field 1) 
is located in the 25 Orionis group (Figure \ref{fig1}; \citealt{briceno05});
it contains 124 TTS, of which 65\% were assigned fibers. 
The second field (Field 2) is located within the Ori OB1b sub-association,
next to the belt star $\epsilon$ Ori (Figure \ref{fig1});
it includes 119 TTS, of which 57\% were assigned fibers.
For each field we obtained three exposures, that were later combined
to aid in cosmic ray removal.
In Table \ref{hcobs} we summarize the characteristics of the observations.

We allocated fibers according to the following priorities:
First, to members already confirmed from low-resolution
spectroscopy, mostly bright (V $\la 17$) members observed with FAST;
we then added some fainter (V$\ga 17$), spectroscopically confirmed members.
Finally we allocated the remaining fibers to other objects in the field 
brighter than V=16, which though not being confirmed members, had
optical or 2MASS colors suggestive of them possibly being candidate young objects.
In each field we observed $\sim 20$ objects with dubious status from
our low-resolution spectroscopy, that could be either WTTS or dMe stars.

To establish the zero point for the radial velocities derived from our
spectra, during the same night 
we obtained template spectra of 14 relatively bright stars with
spectral types similar to our target stars, in a field
located in Selected Area 57, near the north Galactic Pole.
The radial velocities of these stars have been monitored for many years
using the CfA Digital Speedometers \citep{latham92} without showing any
signs of variation, and their absolute velocities have all been established
to an accuracy better than $\rm 0.2\> \kms$ \citep{stefanik99}. 
For wavelength calibration we obtained several Thorium-Argon lamp exposures 
throughout the night;
the thermal shift was monitored to be of a few hundreds of a pixel ($\la 100$ m/s).
Spectra were reduced and extracted using standard IRAF routines.
No sky subtraction was applied. 
To derive the dispersion correction we used 47 lines fitted with a cubic spline.
The RMS fit for the dispersion correction itself was roughly $\sim 100$ m/s, 
however, because of fiber-to-fiber variations, which have not been properly
calibrated out yet, the true scatter is closer to $\pm 0.6\> \kms$.
Typical signal-to-noise (SNR) ratios ranged from $\sim 3$
for the fainter stars (V$\sim 18.5$), to SNR $\sim 100$ for the
brighter objects (V$\sim 13.6$).

Radial velocities were derived using the cross-correlation
routine {\sl xcsao} in the IRAF SAOTDC package.
This routine obtains the radial velocity from the shift of the cross-correlation
peak. 
The R value from the cross-correlation process
is a measurement of the ratio of the peak of the cross-correlation 
to the estimated noise, and is therefore
an indicator of the accuracy of the results \citep{tonry79}.
In our measurements we blocked out H$\alpha$ as well as other emission lines.
For each star in our sample we ran {\sl xcsao} using all the template spectra, we selected
the result with the largest R (the best matching template). 
In our final list we used only $R > 3$ results, which yield velocity
errors $\sigma_e \sim 8 \kms / (1 + R)$ 
(for slowly-rotating stars; \cite[see][]{hartmann86}). In this way we 
avoided spectra contaminated by scattered sunlight, which mostly correspond to
small R values (low SNR) that tend to cluster around the velocity value of
the heliocentric correction.
In order to eliminate
multiple systems in our sample, we selected only objects which showed
a clear single peak in the cross-correlation plot.
The derived radial velocities were corrected for the motion of
the Earth around the Sun.
The final
sample includes 78 "single" TTS with $R > 3$ (47 in Field 1 and 31 in Field 2).

The photometric and spectroscopic measurements 
for these stars are presented in Table \ref{ttsdata}.
The first column provides the CIDA Variability
Survey of Orion (CVSO) running number,
continuing the numbering sequence from \cite{briceno05}; stars
from that article have CVSO numbers $\le 197$. The newly identified young members
of 25 Ori and Orion OB1b presented here start with object CVSO-198.
We provide two values of the Equivalent Width of the H$\alpha$ emission
line ($\rm W[H\alpha]$), 
the first one estimated in our low-resolution spectra, and the second measured in
the Hectochelle spectra. Both values agree for the majority of stars, 
within the measurement uncertainties and allowing for variability of the H$\alpha$ line.
All equivalent widths were obtained using the {\sl splot} routine in IRAF, by
measuring the line respect to the local pseudo-continuum, estimated to be roughly at
the base of the emission feature.

\section{Results and Discussion}\label{results}

\subsection{The 25 Ori population}\label{population}

\begin{figure}
\includegraphics[angle=270,scale=.80]{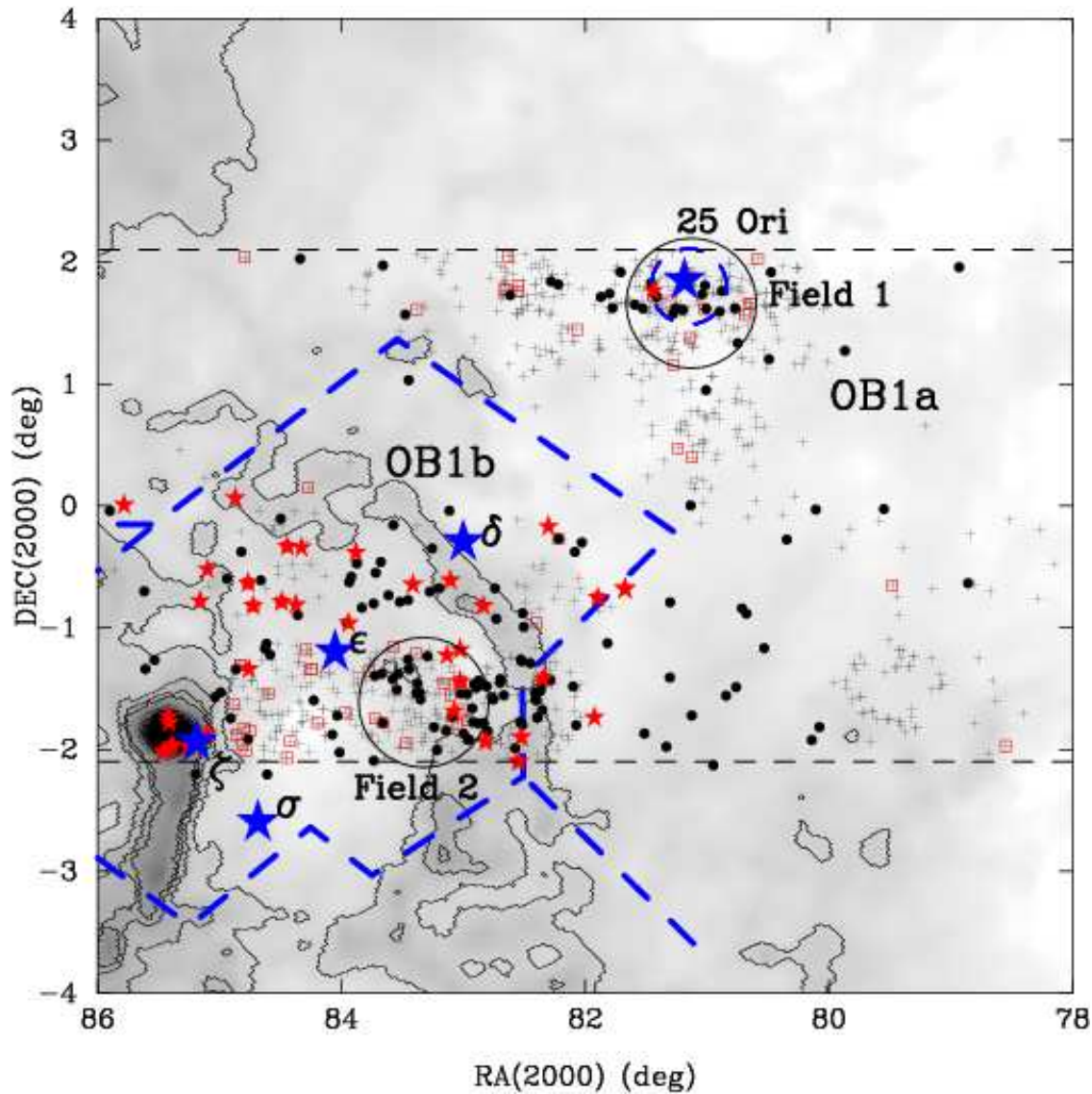}
\caption{Spatial distribution of newly discovered
T Tauri stars in the Orion OB1a and OB1b sub-associations.
Solid dots correspond to the bright (V$\le 16$) WTTS, and small
stars to bright CTTS. Plus symbols indicate
the faint (V$>16$) WTTS, and plus symbols enclosed in open squares the faint CTTS.
The belt stars, 25 Ori and $\sigma$ Ori
are indicated with large starred symbols.
The solid circles indicate the location of the Hectochelle fields
(Field 1 and Field 2).
The smaller dashed-lined circle around 25 Ori corresponds
to the Kharchenko et al. (2005) cluster radius.
The grey scale image shows the dust extinction map from
Schlegel et al. (1998); the countours represent $E(B-V)$
values of 0.1 to 1.5 magnitudes.
The thick dashed line outlines the boundaries between Ori OB1a
and OB1b from \cite{wh77}.
The horizontal dashed lines mark the limits of the
survey region where most of our spectroscopic follow-up
has been carried out so far.\label{fig1}
}
\end{figure}

The 25 Orionis group stands out as a concentration of TTS stars in the Ori OB1a
sub-association (Figure \ref{fig1}), roughly surrounding the B1Vpe star 25 Ori 
($\alpha_{J2000}=81.1867\arcdeg,
\delta_{J2000}=+1.846\arcdeg$).
This star has been classified as a Classical Be object
\citep{yudin01,banerjee00,schuster83}; more recently 
\cite{hernandez05} determined a spectral type B2.
In \cite{briceno05} we already pointed out the existence of
this stellar aggregate as a distinct feature in the spatial distribution
of the brighter (V$\ga 16$) members, albeit with lower number statistics. 
\cite{kharchenko05} classify the region around 25 Ori as a new
cluster (ASCC 16), based on parallaxes, proper motions and V, B-V data
for {\sl Hipparcos} and Tycho-2 stars.
They place the cluster center coordinates
at $\alpha_{J2000}=81.1533\arcdeg,
\delta_{J2000}=+1.800\arcdeg$, $3.4 \arcmin$ south-west of 25 Ori, 
and derive a cluster radius of $0.62\arcdeg$ (Figure \ref{fig1}).
The parallaxes of the {\sl Hipparcos} stars yield a mean distance of
323 pc; we assume here a value of 330 pc \citep[see][]{briceno05}.
We are aware we are missing 
the northernmost region of the 25 Ori cluster, as can be seen in Figure \ref{fig1}. 
However, because our continuing survey ultimately will
span Orion up to $\delta_{J2000}=+6\arcdeg$
we will be in a position to study whether the cluster extends further north.

For the present analysis we defined
a circle of $1^\circ$ radius around 25 Ori as the 25 Orionis group.
Though this is larger than the cluster radius derived by \cite{kharchenko05},
it allows us to improve our member statistics, and at the same time
is not large enough to expect contamination
from the younger stars in the Orion OB1b region.
The \cite{kharchenko05} catalog lists 143 stars within this circle and
south of our $\delta_{J2000}=+2.13\arcdeg$ declination limit (Figure \ref{fig1}):
12 are B-type stars, including 25 Ori,
18 are A-type stars, one of which is the Herbig Ae/Be star V346 Ori
(spectral type A8; \citealt{hernandez05}), and 21 are F-type stars.
A number of these early type stars show IR emission properties consistent 
with membership in this young stellar group. 
\cite{hernandez06} used Spitzer data to look for circumstellar disks 
in a subset of the \cite{kharchenko05} stars.
They studied 7 of the 12 {\sl Hipparcos} B-type stars,
and found one with 24$\mu$m excess emission and lack of near-IR excesses indicative of
a debris disk. Their sample also included 13 A-type stars, out of which 7 had debris
disks, and 1 (V346 Ori) had an optically thick disk, for a disk fraction among A-stars of 62\%;
of the 5 F-type stars in their sample 3 (60\%) had debris disks. In our
subsequent discussion we adopt these sources
as early type members of the 25 Ori group.

If a standard Initial Mass Function (IMF; e.g. \citealt{kroupa2001}) is assumed, 
one would expect that a cluster formed with 12 B-type stars will
produce roughly 12 A-stars, 12 F-stars, 18-G stars, 48-K stars and 336-M stars
(down to $\sim $M6).
Within this area we identify 197 TTS from our low-resolution follow up spectroscopy
(124 inside our 25 Ori Hectochelle field - Field 1),
all of them K and M-type; 21 objects were reported in \cite{briceno05}
and 176 are new identifications, whose detailed properties will be
discussed in \cite{briceno07}.
The magnitude limit of our spectroscopic follow-up effectively places our
completeness at a spectral type of roughly M4 (we discuss this with better
number statistics in \citealt{briceno07}); therefore, our sample is
missing a number of the later type M-stars.
Still, the number of low-mass young stars we have identified within this region
is in agreement, within a factor of $\la 2$, to expectations from
a typical IMF.

\subsection{Kinematics}\label{kinematics}

\begin{figure}
\plottwo{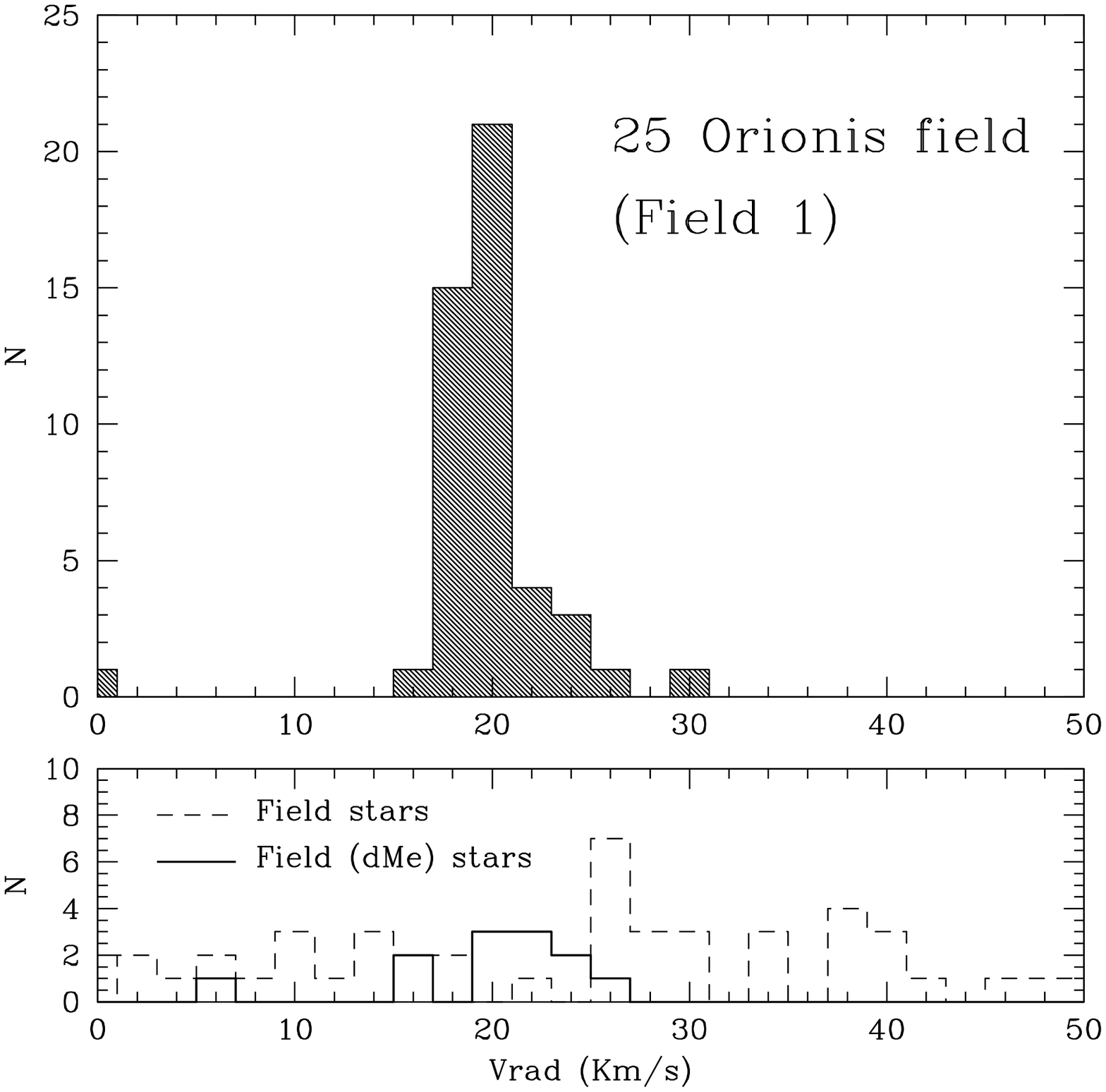}{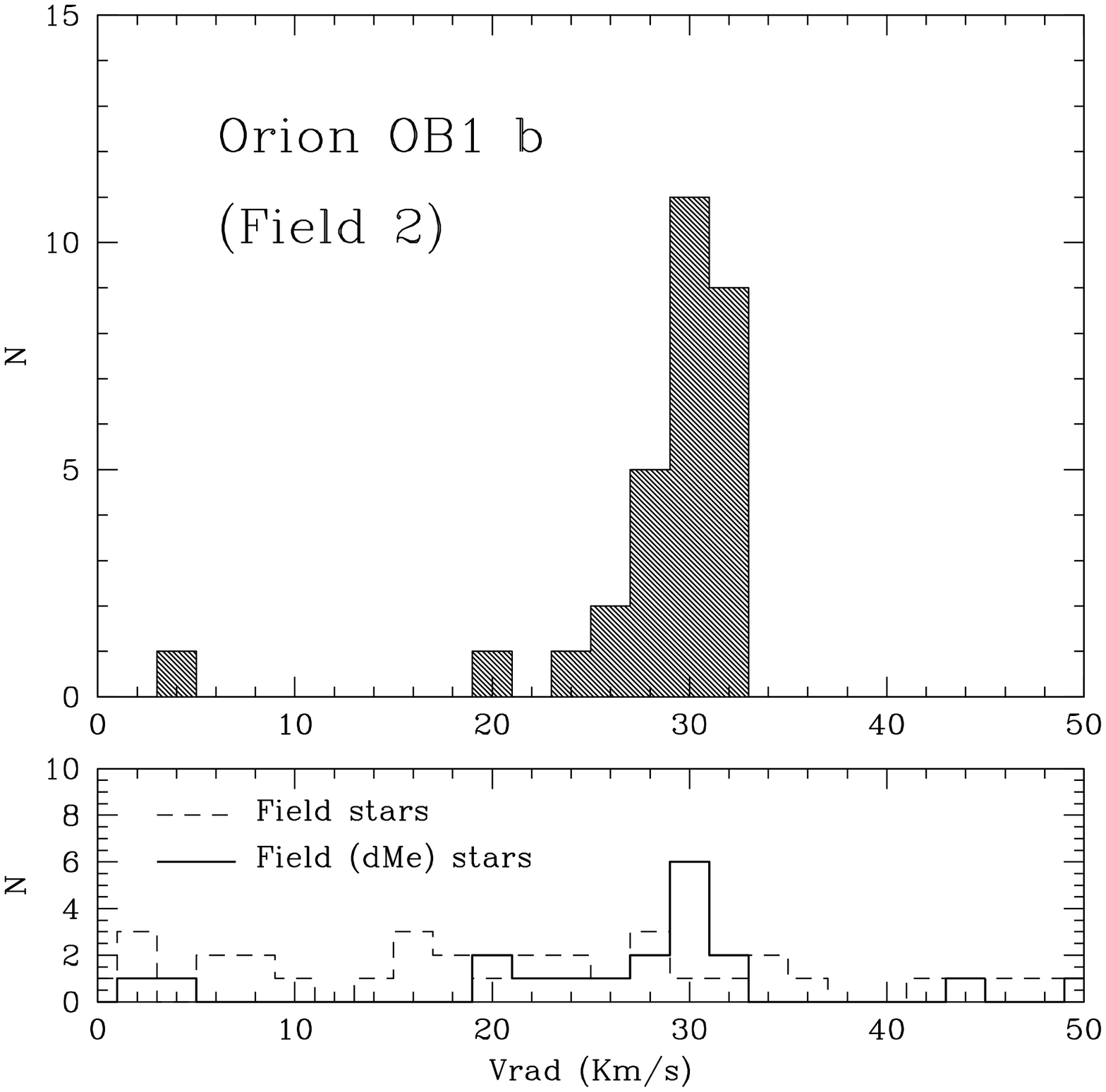}
\caption{Heliocentric radial velocities in our 25 Ori and OB1b Hectochelle fields.
In both plots
the solid histogram represents stars classified as members from our
low-resolution spectra. Velocity bins are $2 \kms$ wide. 
The kinematics of the TTS in each region are characterized by strong and narrow peaks
at distinct velocities.
The lower panels show the distribution of radial velocities for stars
classified as field objects because of their very strong H$\alpha$ absorption
in our Hectochelle spectra, and stars classified as dMe in
our low-resolution spectra.\label{fig2}
}
\end{figure}

Figure \ref{fig2} shows the velocity distribution of stars in Table \ref{ttsdata}.
Because in our sample $\rm R(median)=8$ 
the typical velocity errors are $\sim 1 \kms$.
The two regions show clearly distinct
kinematics, with well defined peaks located at 
$19.7 \pm 1.7 \kms$ for the 25 Ori field and $30.1 \pm 1.9 \kms$ for the Ori OB1b
field.
The narrow cores $\sim \pm 2 \kms$ of both distributions are
not much larger than the estimated errors, and consistent with
velocity dispersions in typical regions of star formation.
The peak of the velocity distribution in Ori OB1a is consistent with
the majority of radial velocity measurements for 25 Ori
($18.9 \pm 2\kms$, \citealt{frost26}; $19.9 \pm 2\kms$, \citealt{plaskett31}; 
$19.3 \pm 2\kms$, \citealt{wilson53}; $24 \pm 4 \kms$, \citealt{morrell91}), 
suggesting that the bulk of the TTS are indeed related to the star 25 Ori.
The presence of a $10 \kms$ shift between the cores of the velocity distributions
in OB1a/25 Ori region and  OB1b qualitatively supports the findings that these
two sub-associations are at very different distances, though nearly along the
line of sight \citep{wh77,wh78,brown94,hernandez05}.

In a recent study, \cite{jeffries06} found two distinct velocity components
when analyzing velocity data toward four fields around $\sigma$ Ori, located
in the sub-association Ori OB1b.
Their kinematic {\sl group 1} has velocities $20 < V_r < 27 \kms$.
Taking the average value of $23.8 \pm 0.7 \kms$  
determined by \cite{morrell91} for Ori OB1a,\footnote{The errors
quoted by \cite{morrell91} in their Table 5, for the average $V_r$
derived for each Ori OB1 sub-association are the error of the mean,
$\sigma(V_r) / \sqrt N$. From their OB1a and OB1b measurements, the
actual mean internal error is $3.9 \kms$
and their mean external error is $14 \kms$.} 
they relate stars in {\sl group 1} with this sub-association; this is consistent
with suggestions that the closer Ori OB1a region in fact
extends in front of Ori OB1b \citep{briceno05,sherry04}.
The velocities of {\sl group 1}
roughly overlap our distribution for 25 Ori. However, it is
interesting to highlight that the peak of the distribution for the {\sl group 2}
of \cite{jeffries06} is located at $\sim 24 \kms$, $4 \kms$ off 
the $\sim 20 \kms$ peak of our 25 Ori $V_r$ distribution. With errors in both
studies in the range $0.5 - 0.9 \kms$, this is a statistically robust difference.
If the kinematics of the "field" population of Ori OB1a is indeed characterized by
the velocity distribution of {\sl group 1} in the \cite{jeffries06} sample, then
our results strongly argue in favor of 25 Ori cluster being a 
a distinct entity in velocity space, different from the general population of OB1a,  
which is widely spread with a low spatial density
\citep[c.f. Figs. 5 and 6 of][]{briceno05}.
The distribution of velocities in 25 Ori (Field 1) looks somewhat
skewed to higher velocities ($V_r \ga 22 \kms$; Figure \ref{fig2}), though with
few objects. This is what would be expected from the presence of a few
interloper PMS stars from the widespread population of OB1a. 
If this degree of contamination (5/47) is taken as representative of the
field PMS population of Ori OB1a in the 25 Ori area, we would expect that
of the 124 TTS within our Hectochelle field, $\sim 13$ could be "contaminants"
from OB1a.

The OB1b stars in our sample lie to the west of the main distribution of
molecular gas, corresponding to the respective ends of the Orion A and
B clouds (Figure \ref{fig1}).  The gas nearest the stars has a heliocentric radial
velocity $\sim 29 \kms$ \citep[e.g., Figure 5 of][]{wilson05},
velocity as the stars in our Hectochelle field. 
There is also a degree of asymmetry
in the velocity distribution for Ori OB1b (right panel in Figure \ref{fig2}), 
this time extending to lower $V_r$, 
toward the same velocity range expected of OB1a stars and of 25 Ori. 
This is not surprising if we recall that Ori OB1a is closer than OB1b, 
and much more spatially extended, such that
along any line of sight in the direction of OB1b
there probably will be some PMS members of the general population of Ori OB1a,
that will contribute at these velocities.
Another fact worth pointing out is that the gas is
present over a range of several $\kms$, which could also account for this asymmetry.
The Orion belt star nearest to our Hectochelle Field 2, in Ori OB1b, is $\epsilon$ Ori,
with estimated velocities that range from $29 \pm 2 \kms$ \citep{frost26} to 
$23 \pm 9 \kms$ \citep{morrell91}, 
though most measurements cluster around $\sim 27 \kms$.
\cite{brown94} classify $\epsilon$ Ori as a member of OB1b. The {\sl Hipparcos}
parallax is $2.94 \> mas$ \citep{perryman97}, which corresponds to a distance 
of 412 pc, consistent with our assumed distance of 440 pc.
The TTS in our southern Hectochelle field share the same velocity 
as the nearby gas and $\epsilon$ Ori, indicating that these objects
indeed belong to what is called the OB1b sub-association.

In their study \cite{jeffries06} found for their
{\sl group 2} velocities in the range $27 < V_r < 35 \kms$, peaking at $\sim 31 \kms$. 
They identify this group as the "$\sigma$ Ori cluster", with an age of $\sim 3$ Myr and at a
distance of 440 pc, similar to Ori OB1b; however, they argue it is
kinematically distinct to OB1b, because they assume the value of 
$V_r = 23.1 \pm 1.4 \kms$ from \cite{morrell91} for this sub-association.
Our findings do not support their claim. First, the $\sim 30 \kms$ peak velocity we
find for OB1b, and the small width of the distribution shown in Figure \ref{fig2},
strongly suggest that OB1b and the $\sigma$ Ori cluster share similar kinematics.
\footnote{Despite having similar kinematics, \cite{hernandez07} found
that $\sigma$ Ori appears to be younger than the general population of Ori OB1b,
and with a higher disk fraction (their Figure 13).}
Second, if we recompute the average group velocity for the OB1b stars in 
\cite{morrell91}, but considering only objects with reliable membership,
that are not spectroscopic binaries, and flagged as having 
constant $V_r$, we find $V_r = 23.7 \kms$ with a $1\sigma$
dispersion of $\pm 9 \kms$. Therefore, the peak velocity of the \cite{jeffries06}
{\sl group 2} is consistent with the radial velocity of both our sample of 
spectroscopically confirmed low-mass members, and samples of early type stars 
in Ori OB1b.

In contrast to the strongly peaked velocity distributions for the young
stars, field stars (objects showing strong H$\alpha$ absorption in our
Hectochelle spectra) exhibit an almost flat distribution, with
stars at all velocity bins (lower panels in Figure \ref{fig2}). 
Other objects recognized as part of the
field population are dMe stars, which look very similar to WTTS in that
they have weak H$\alpha$ emission, but lack the strong Li~ I 6707{\AA }
absorption feature characteristic of M stars younger than $\sim 15-20$ Myr.
However, the velocity distribution of dMe stars in both fields shows a small
excess of objects at the velocity peak for each region.
A few of these objects could actually be WTTS, 
but the SNR of their low resolution spectra does 
not allow detection of the Li~ I line.
But even if this were the case, we expect to be missing no more than 1-2 objects
in 25 Ori and 2-4 in Orion OB1b.

\subsection{Lithium}\label{lithium}

\begin{figure}
\includegraphics[angle=0,scale=.75]{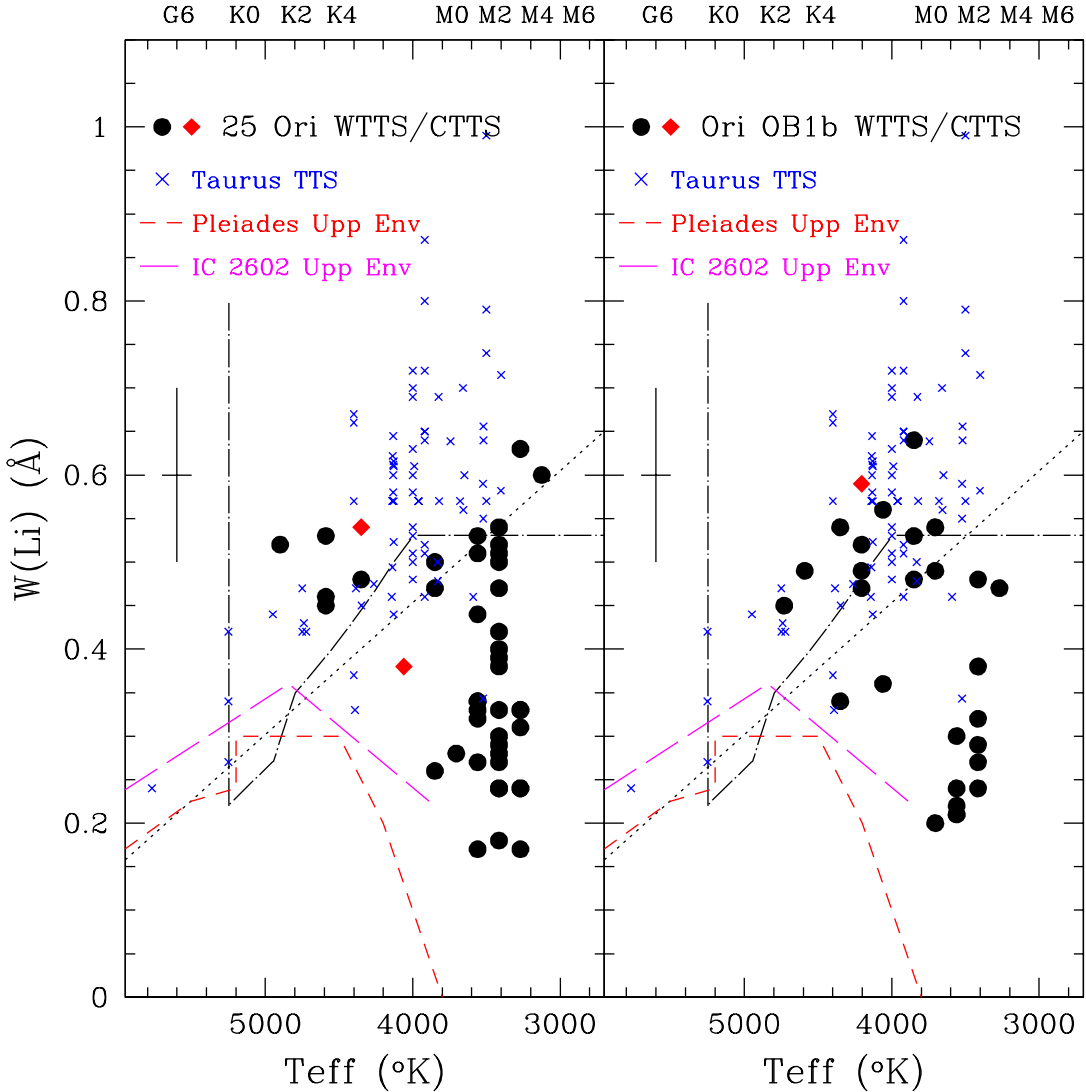}
\caption{Equivalent width of the Li~ I 6707{\AA } line, as measured in
our low-resolution spectra, plotted against the effective temperature
derived for each star. Left panel: stars in our 25 Ori Hectochelle field (Field 1).
Right panel: stars in the OB1b Hectochelle field (Field 2).
WTTS are shown as dots and CTTS as diamonds.
For clarity, we have plotted here our original
$\rm W(Li~I)$ measurements. These values have later been rounded off
to one decimal place for their presentations in Table \ref{ttsdata}, 
more representative of our actual error.
Data for Taurus PMS stars (age $\sim 1-2$ Myr) are shown as $\times$'s 
\citep{basri91,martin94}.
The dotted line traces the lower boundary to the majority of Taurus TTS.
The long dash-dot line corresponds to the lithium isoabundance line 
from \cite{martin97}, which separates WTTS from PTTS.
We show as a short-dash line the upper envelope for the Pleiades cluster
(age $\sim 125$ Myr; \citealt{stauffer98}. 
Data from \citealt{soderblom93} and \citealt{garcialopez94}),
and in the long-dash line the upper envelope of the IC 2602 cluster 
(age $\sim 30$ Myr; \citealt{stauffer97}. Data from \citealt{randich97}).
Our typical error bar is also indicated in each panel.\label{fig3}
}
\end{figure}

As mentioned in \S \ref{lowres} and above, the presence of the Li~ I 6707{\AA } line strongly in
absorption is a clear indicator of youth in K and M-type stars. In Figure \ref{fig3},
we show the distribution of Li~ I equivalent widths, $\rm W(Li~ I)$, 
measured in our low-resolution spectra, plotted as a function of the
effective temperature for each star, derived from the adopted spectral type and
the temperature scale in Table A5 of \cite{kh95}.
In order to compare with our sample, 
we show the lithium equivalent width measurements by \cite{basri91}
and \cite{martin94} of Taurus TTS, representative of $\sim 1-2$ Myr
old, low-mass stars. 
The dotted line indicates the lower boundary of W(Li~I) 
for the majority of Taurus TTS.
We also plot the upper envelopes for the Pleiades cluster
(age $\sim 125$ Myr; \citealt{stauffer98}) 
using data from \cite{soderblom93} and \cite{garcialopez94},
and for the IC 2602 cluster (age $\sim 30$ Myr; \citealt{stauffer97}),
from the dataset of \cite{randich97}.
All the 25 Ori and Ori OB1b members fall above the upper cluster
envelopes; however, an important fraction of the M-stars in each field
fall below the lower boundary of the Taurus TTS. Also, in both
Orion fields the observed $\rm W(Li~ I)_{max} \sim 0.6$ is lower than in Taurus.
The large spread toward lower values of $\rm W(Li~ I)$ for the M-type stars
has also been observed by \cite{kenyon2005} in $\sigma$ Ori.
Comparison of the $\rm W(Li~ I)$ in 25 Ori and OB1b members
shows that while 40\% of all the Ori OB1b members in
Table \ref{ttsdata} fall within the Taurus TTS locus
only 22\% of the TTS in the 25 Ori group
share this region of the diagram (see Figure \ref{fig3}) . 
This analysis includes all stars with velocities
within $\pm 3\sigma$ of the peak $V_r$ in each field.
In Figure \ref{fig3} we have also plotted the line separating
WTTS from post-T Tauri stars (PTTS). \cite{herbig78} used the term
PTTS for PMS stars which have lost their T Tauri properties;
\cite{martin97} identify as PTTS those PMS stars with essentially WTTS
properties, except they fall below the
the lithium isoabundance line plotted in Figure \ref{fig3}. 
In general, PTTS are expected to be slightly more evolved TTS.
In the Ori OB1b field, the WTTS/PTTS fraction is 0.48;
in the 25 Ori field it is 0.25. In contrast, the WTTS/PTTS fraction 
for Taurus members is 5.57.
We argue that the lower $\rm W(Li~ I)_{max}$,
the increasing fraction of TTS with $\rm W(Li~ I)$ values below the Taurus lower
boundary as a function of age, and the decreasing WTTS/PTTS fraction,
suggest we are looking at
the effects of Li depletion over the age range spanned by these three regions.

\subsection{Ages}\label{ages}

\begin{figure}
\plottwo{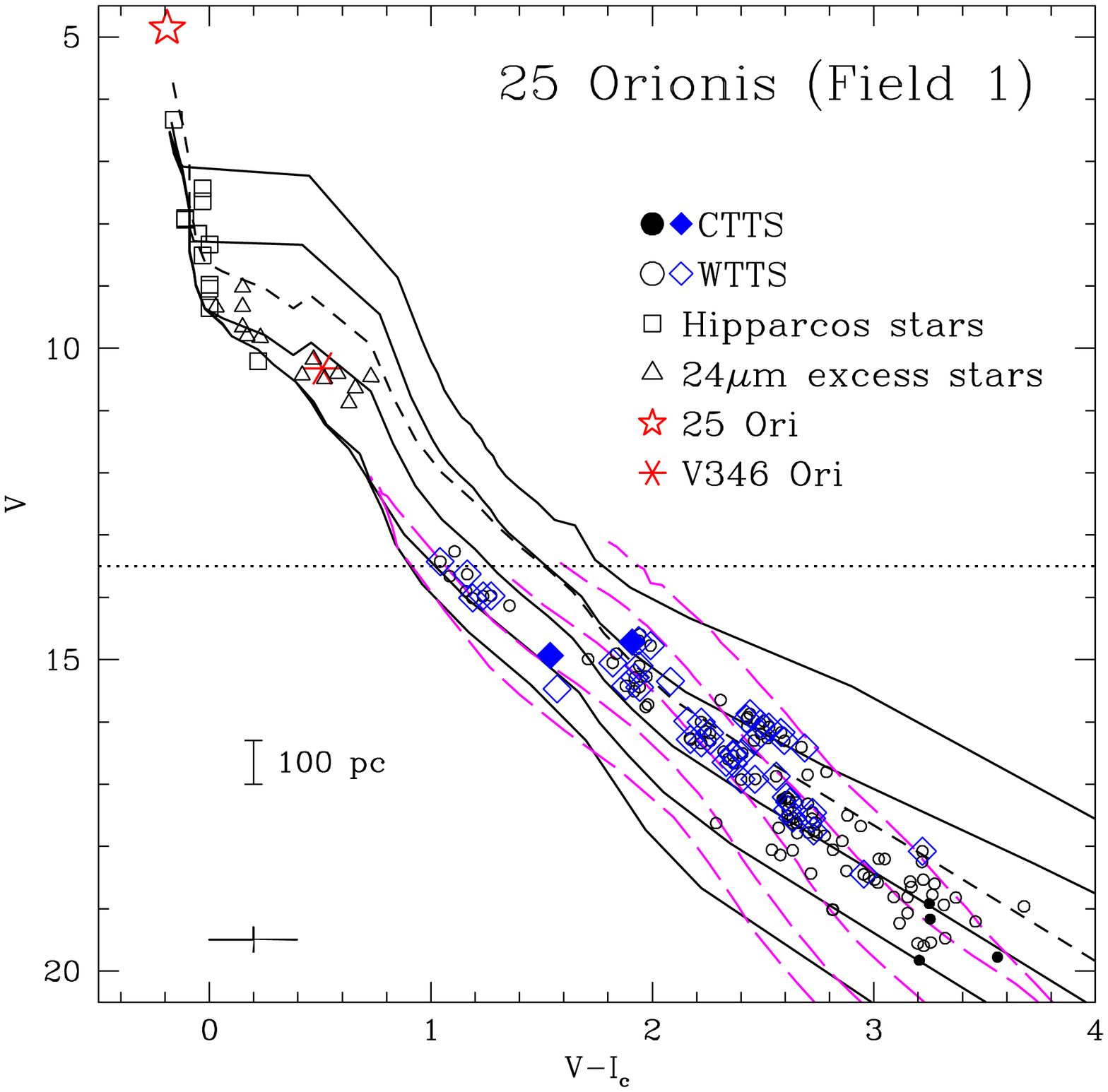}{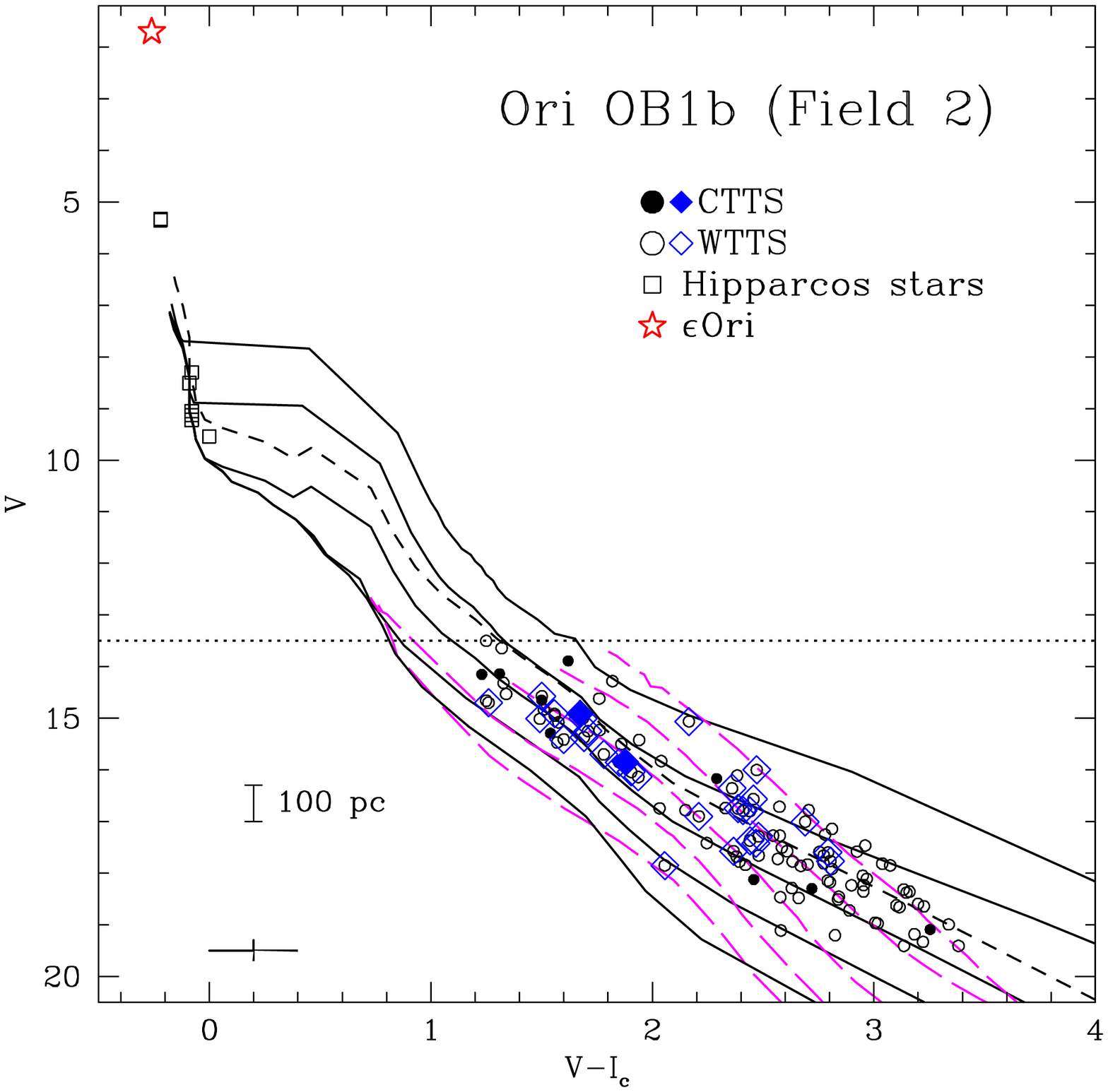}
\caption{Observed color-magnitude diagram of {\sl all} young stars in our
Hectochelle fields.
The T Tauri stars \citep{briceno05,briceno07} are indicated with small open circles
(WTTS) and solid dots (CTTS). Stars observed with Hectochelle
(Table \ref{ttsdata}) are indicated with large open diamonds (WTTS) and solid
diamonds (CTTS).
The B type {\sl Hipparcos} stars within each area are shown as open squares.
For the 25 Ori CMD we included other early type stars (A-F) showing IR 
excess emission at 24$\mu$m \citep[][; open triangles]{hernandez05}.
We plot two sets of iscohrones, Siess et al. (2000; solid lines) and
Baraffe et al. (1998; long dashed lines), from top to bottom,
 1, 3, 10, 30 and 100 Myr (which we adopt as the zero-age main sequence).
The short dashed line in the left panel
indicates the 0.75 magnitude offset, for the \cite{siess00}
10 Myr isochhrone, expected from unresolved binaries. 
The error bar at the lower left indicates the
typical uncertainty at our faint magnitude limit.
The vertical bar indicates the magnitude shift corresponding to 100 pc,
roughly the difference between Orion OB1a and OB1b.
The horizontal dotted line indicates the typical saturation limit of our photometry.\label{fig4}
}
\end{figure}

In Figure \ref{fig4} we show the color-magnitude diagrams (CMDs), 
of {\sl all} TTS
within our two Hectochelle fields (open and solid circles), 
and for the subset of members for which 
high-resolution spectra were obtained (HC-only, 
listed in Table \ref{ttsdata}; open and solid diamonds).
As in \cite{briceno05} we adopted distances of 330 pc for the 25 Ori
group and 440 pc for the Ori OB1b field.
We also show two sets of isochrones for 1, 3, 30, and 100 Myr from
\cite{baraffe98} and \cite{baraffe98}.

The TTS in the 25 Ori field follow a relatively well defined band
in the CMD.
The spread observed is largely consistent with the upper
limit of 0.75 magnitudes expected from unresolved binaries
(the dashed-lined isochrone in Figure \ref{fig4}), plus a small spread introduced
if we assume the stars follow some distance distribution along the line of sight 
($1 \arcdeg$ at 400 pc corresponds to $\sim 7$ pc, or 0.04 magnitudes). 
Variability should not affect in an important way our CMDs because we
use for each object the mean magnitudes derived from our multi-epoch
optical survey \citep[see][]{briceno05}.
From our photometry and spectral types for
the TTS members, assuming the intrinsic colors in Table A5 of \cite{kh95},
and the extinction law of \cite{cardelli89} with $\rm R_V=3.1$,
we compute an overall mean $\rm A_V=0.29$ mags for the 25 Ori cluster, 
in good agreement with the E(B-V)= 0.09 obtained by \cite{kharchenko05}.
For the 25 Ori group we also plot the V and $\rm V-I_C$ values for {\sl Hipparcos} stars 
\citep[including 25 Ori;][]{perryman97}, V346 Ori and the 
excess IR emission early type stars from \cite{hernandez06};
the V, B photometry for these later objects is from \cite{kharchenko05},
transformed to $\rm V-I_C$ assuming a mean $\rm A_V=0.29$ and the appropriate
colors for their spectral types from \cite{kh95}.
Similarly, we show the location of the {\sl Hipparcos} B-type stars and the belt star
$\epsilon$ Ori in the CMD diagram of Ori OB1b.
All these early type stars 
fall on or close to the the ZAMS, consistent with them
being the higher mass members of these stellar populations.

The band of TTS in the 25 Ori field is
roughly encompassed by the 3-10 Myr isochrones.
At the bright end of our TTS sample (V$\la 14$) the stars seem to cross
over to the left, towards the 30 Myr isochrone; this could be due to
the photometry for some of these stars being affected by
saturation (the bright limit for our photometry varies with each
CCD in the QuEST camera, but is roughly between V$\sim 13-14$).
Alternatively, this could be due to uncertainties in the
location of the birthline at these range of masses \citep{hartmann01}.

At the faint end we note a deviation from the theoretical
models that we attribute to limitations in the currently available 
PMS isochrones. At V$\ga 17.5$, $\rm V-I_C \ga 2.5$ the bulk of our stars 
in the CMD seems to deviate slightly downwards from the \cite{siess00}
isochrones. 
With the Baraffe tracks,
which are considered to be more appropriate for low mass
stars \citep[e.g.][]{briceno02,luhman03}, the behavior is opposite, 
the lower mass stars seem systematically younger
(or the tracks bend down too steeply). 
A similar discrepancy is found in the Ori Ob1b field.
Overall, the Baraffe models 
produce mean ages with are $\sim 20-30\%$ younger than those derived
from the the Siess isochrones.
For the larger sample we derive an average age of
$\sim 10$ Myr from the \cite{siess00} isochrones, and of $\sim 7$ Myr from the
\cite{baraffe98} isochrones, in agreement with our determination in
\cite{briceno05} with smaller number of objects.
For the HC-only sample in 25 Ori the resulting average ages are similar.
For both TTS samples within the Orion OB1b field, the resulting average ages
are $\sim 5$ Myr from the Siess et al. isochrones and $\sim 4$ Myr
from the Baraffe et al. models. In the OB1b field 
we find an average $A_V=0.38$ mags.

Within the uncertainties, the age derived
here for 25 Ori is the same as for the general Ori OB1a sub-association
in \cite{briceno01,briceno05}, though for this larger region the spread in ages
for the individual stars is larger, 
probably caused in some measure by the increase in distance depth along
the line of sight as one encompasses much wider areas.
With a well defined, spatially confined, numerous sample in Ori OB1b, 
resulting in a smaller spread in the CMD, the age difference between
this younger region and the older Ori OB1a is now firmly established.
This age difference of a factor $\sim \times 2$ 
is consistent with the indications of Li~I depletion we discuss 
in \S \ref{lithium}, and with the decay in accretion indicators
shown by \cite{briceno01,briceno05} and considered here in 
the following section.\\
Our finding of slightly different kinematics between the 25 Ori cluster
and what can be interpreted as the widely spread PMS low-mass population
of Ori OB1a, suggest that although both stellar populations probably
share a common origin, a previously unrecognized degree of substructure
is present in this region.

\subsection{Accretion indicators}\label{accretion}

The fraction of CTTS, or stars thought to be accreting from their circumstellar disks, 
can be derived by comparing the value of W(H$\alpha$) for each spectral type
in our low resolution spectroscopy, with the criteria in \cite{wba03}:
a star is considered to be a CTTS if $\rm W(H\alpha) \ge 3$ in the range K0-K5,
$\rm W(H\alpha) \ge 10$ for K7-M2.5,  $\rm W(H\alpha) \ge 20$ for M3-M5.5,
and  $\rm W(H\alpha) \ge 40$ for M6-M7.5.
Inside each 0.5 deg radius Hectochelle field, and considering {\sl all} TTS,
the CTTS fraction in the 25 Ori field is 5.6\% (7 out of 124 stars),
while in the Ori OB1b field the CTTS fraction is 12.6\% (15 out of 119 stars, 
$\sim 2\times $ higher).
These values are almost a factor of $\sim 2$ lower than those reported in
\cite{briceno05}. However, this apparent discrepancy can be easily explained 
by considering the following: first, in \cite{briceno05} our sample for Ori OB1b included 
regions like the very young NGC 2024 cluster ($\la 1$ Myr) which has a large number of CTTS, 
and is nominally located within Ori OB1b if one uses the \cite{wh77} boundaries. 
Second, as one increases the census of PMS stars in older regions like Ori OB1a, the
most frequent type of members are WTTS, which tends to lower the accretor fraction. 
If we assume that our present CTTS fraction in 25 Ori is more representative of this
region, then we find it is a factor of $2.5 - 3\times $
larger than what \cite{sicilia05} obtained in the $\sim 10$ Myr old NGC 7160 (
1 accretor in their sample of 55 members), but much smaller than the $\sim 40$\% they
report for the $\sim 4$ Myr old cluster Tr 37.
The higher CTTS fraction in 25 Ori when compared to NGC 7160
could be because this region is indeed younger,
or due to the small number statistics in the \cite{sicilia05} sample.

\subsection{Spatial Distribution}\label{spatial}

\begin{figure}
\includegraphics[angle=270,scale=.80]{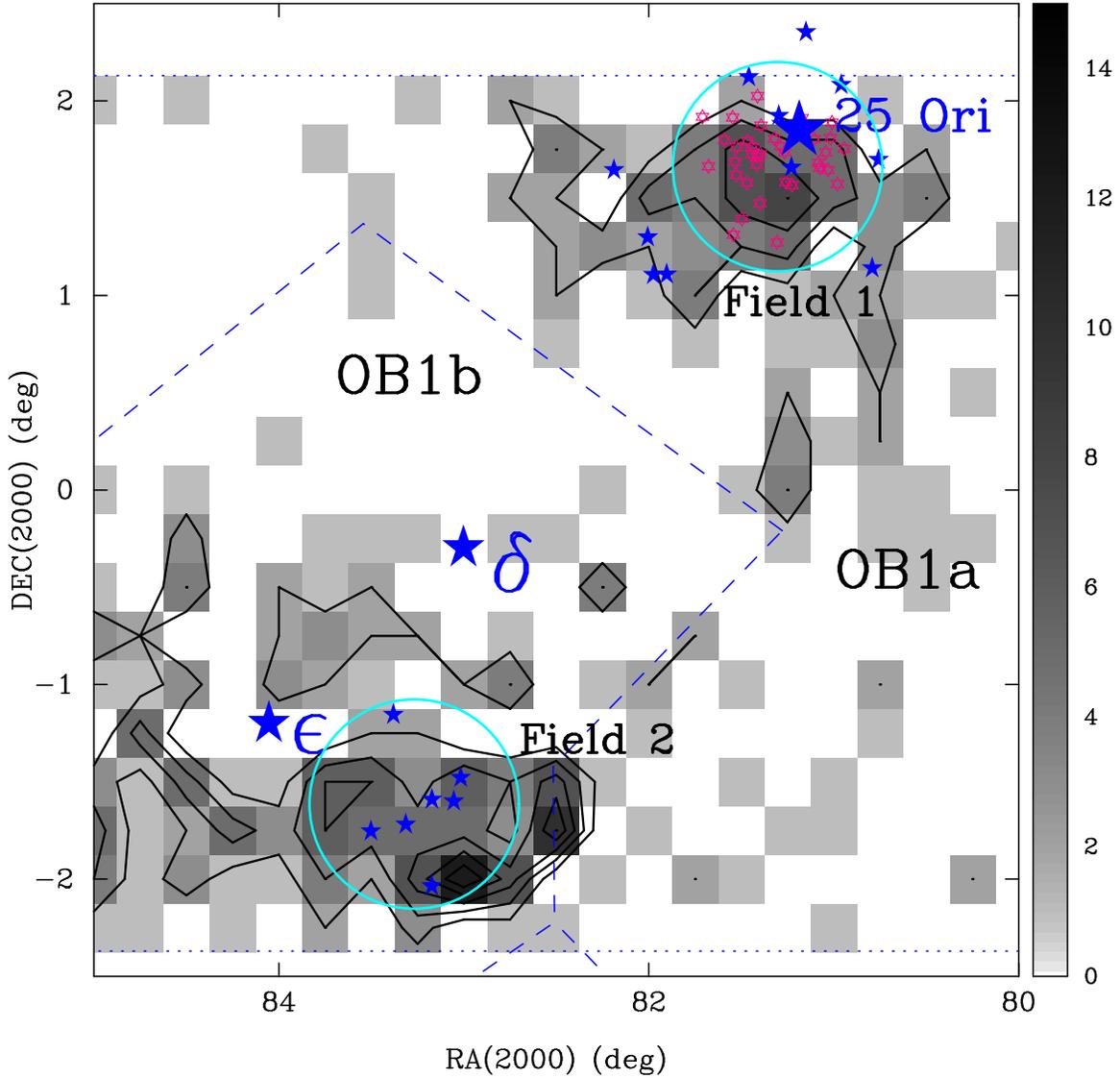}
\caption{Spatial density map of bright (V $\le 16$) T Tauri stars
in the Orion OB1a and OB1b sub-associations. The contours
correspond to stellar densities of 2, 4, 6, 8, and 
10 stars per 0.25 deg square bin.
The largest starred symbol marks the location of 25 Ori
($\alpha_{J2000}=81.1867\arcdeg,
\delta_{J2000}=+1.846\arcdeg$).
The smaller stars around 25 Ori indicate the B-type stars from
the list of Kharchenko et al. (2005) in their ASCC 16 cluster.
The Orion belt stars $\delta$ and $\epsilon$, in the Ori OB1b region,
are also indicated with large starred symbols. The short dashed
line around the belt stars indicates the Ori OB1b boundaries of
\cite{wh77}.
The two light circles show the 1 deg wide Hectochelle fields. 
The small open stars in the 25 Ori Hectochelle field indicate
TTS with velocities within 1$\sigma$ of the $V_r$ peak for the
cluster ($V_r(peak) = 19.7 \pm 1.7\kms$).\label{fig5}
}
\end{figure}

In order to provide a first look at the spatial distribution of young stars in this
aggregate, we computed a spatial density map by counting the number of
TTS in squares $0.25 \arcdeg$ in side; because the spectroscopic followup is
most spatially uniform for the brighter stars, we only considered TTS with V$\le 16$
(the non-overlapping Hectospec circular fields produce circular ``imprints'' on
the spatial distribution for the fainter members, as can be seen in Figure \ref{fig1}).
The result is shown in Figure \ref{fig5}. The largest starred symbol marks the location
of 25 Ori, and the belt stars $\delta$ and $\epsilon$ Ori;
the small solid starred symbols correspond to B-type stars included as
probable members of the 25 Ori cluster in the list of \cite{kharchenko05},
and {\sl Hipparcos} B-stars classified as members of Ori OB1b by \cite{hernandez05}.
There is a peak in the spatial distribution of low-mass young stars
in the 25 Ori region, containing 8 TTS (an implied density of 128 $stars/deg^2$),
at $\alpha_{J2000}=81.3\arcdeg, \delta_{J2000}=+1.5\arcdeg$,
$23.6 \arcmin$ south-east of 25 Ori. 
\cite{kharchenko05} place their cluster center at $\alpha_{J2000} = 81.15 \arcdeg$,
$\delta_{J2000}= +1.80 \arcdeg$, which given our bin size agrees with our
determination.
The density of stars falls off to $\la 1$/bin
at a radius of $\sim 1.2\arcdeg$, which corresponds to 7 pc at the assumed
distance of 330 pc. This value is slightly larger than what \cite{sherry04} found
for the younger $\sigma$ Ori cluster ($\sim 3-5$ pc). At a velocity dispersion
of $\sim 1 \kms$, an unbound stellar aggregate would expand roughly 1 pc every 1 Myr.
If the 25 Ori group resembled the $\sigma$ Ori cluster at an age of $\sim 4$ Myr,
a naive dynamical picture would have it evolve to a cluster radius of $\sim 7-9$ pc
at $\sim 8$ Myr.
However, we must caution that the detailed spatial structure cannot yet be fully
investigated from our present data.
First, because of gaps between
the CCD detectors in the QuEST CCD Mosaic Camera, artifacts appear
such as the E-W gap apparent at $\delta= +1.70\arcdeg$ (Figure \ref{fig1}).
Second, we have not yet completed follow-up spectroscopy of the photometric
candidates located north of $\delta_{J2000}=+2.13\arcdeg$ (Figures \ref{fig1} and \ref{fig2}). 
Therefore, if more young stars turn up northward of that line,
the actual membership and spatial extent of
the 25 Ori group may be larger than shown here, and the point with the
highest density of members could well shift somewhat.
To illustrate this point we have plotted as small magenta stars
the TTS with $V_r$ within 1$\sigma$ of the peak velocity for the 25 Ori cluster
(34 objects).
It can be seen that these stars concentrate in the northern half of the
Hectochelle field, with 74\% located in this area.
The highest density is in the NE quadrant of the field, mostly
within the highest contour of the spatial density map.
We conclude that the spatial distribution of the kinematical members
within the Hectochelle field suggests the cluster may extend further north
(the apparent dearth of members close at the northernmost part of the field
could well be an incompleteness effect).
\cite{mcgehee06} used Sloan Digital Sky Survey Data (SDSS) to look for TTS
in a $2.5\arcdeg$ wide equatorial strip between $\alpha_{J2000} = 75-90 \arcdeg$ and 
$\delta_{J2000}= -1.25\arcdeg - +1.25\arcdeg$. In his Fig.8 he shows an excess of 
candidate PMS stars at $\alpha_{J2000} \sim 81\arcdeg$
and $\delta_{J2000}= +0.8 - +0.5\arcdeg$, which he identifies as the south tip of
the 25 Ori cluster. Because of this he derives a larger cluster radius of 8-11 pc.
Though his sample does not overlap the 25 Ori
nor the Ori OB1b fields discussed here, we can compare with the larger scale
distribution of stars in our Orion OB1 survey.
The feature identified by McGehee as part of 25 Ori
can be seen as an elongated density enhancement in our spatial 
density map, roughly $1.2\arcdeg$ south of 25 Ori; additionally, 
we find a number of faint TTS at this location
(Figure \ref{fig1}). Even taking into account this group of young objects, 
there is an apparent paucity of stars beyond $\sim 1.2\arcdeg$ south from 25 Ori;
radial velocities need to be obtained before these 
objects can be interpreted as related to the 25 Ori cluster.

At present we are obtaining additional Hectochelle fields north and south 
of the 25 Ori field discussed here, and also in nearby regions of Ori OB1a. 
These observations
will not only help us constrain the spatial extent and refine the kinematics
of this unique sample of young, low-mass stars, but will also provide for the
first time important information on the actual velocity distribution of the 
sparse PMS population of the Orion OB1a sub-association.

\section{Summary and Conclusions}\label{conclusions}

We have used the Hectochelle spectrograph on the 6.5m MMT to obtain high 
resolution spectra of 149 low-mass PMS stars distributed in two fields
located in the Orion OB1 association, one field placed on the 25 Ori group
originally identified by \cite{briceno05}, the other near
the belt star $\epsilon$ Ori, in the Ori OB1b sub-association.

We determined radial velocities with errors $\la 1 \kms$ for 78 stars
(47 in the 25 Ori field and 31 in the Ori OB1b field).
The PMS members of the 25 Ori group have radial velocities 
in a narrow range, peaking at $19.7 \kms$;
this value is very close to that of 25 Ori itself, suggesting this may be
indeed the most massive star in this stellar aggregate.
The stars belonging to the Ori OB1b sub-association peak at
$30 \kms$, and also show a very narrow distribution.
These two differing values clearly show that these are two separate populations, 
and give strength to findings of a significant distance difference between these two groups.
However, the 25 Ori group not only is distinct in velocity space from the Ori OB1b
sub-association.  The cluster peak $V_r$ is $4 \kms$
off from the peak velocity found by \cite{jeffries06} for stars they
identify as belonging to the Orion OB1a sub-association.  Therefore the 25 Ori group
constitutes a distinct kinematic group, clearly differentiated in velocity space
from both the younger and more distant OB1b, and from what seems
to be the general, widely spread population of Ori OB1a.

The color-magnitude diagrams yield ages of 7-10 Myr for the 25 Ori group,
and of 4-5 Myr for Ori OB1, depending on the assumed set of isochrones. 
The observed spread in each sample is roughly consistent with the upper limit
expected for a population of unresolved binaries, combined with the distance
depth estimated for each region.
Our finding of a similar age but
slightly different kinematics between the 25 Ori cluster
and what can be interpreted as the widely spread PMS low-mass population
of Ori OB1a, suggest that both stellar populations probably
share a common origin, but also that a
previously unrecognized degree of spatial and kinematic substructure
is present in this region.
The overall behavior of the Li~I equivalent widths from one region to
the other, and the decline of a factor $\sim 2$  in the fractions of CTTS 
are consistent with the 25 Ori group being significantly older than Ori OB1b,
independent of the absolute age calibration.

A first approach to the spatial structure of the 25 Orionis group, with
the caveat of an incomplete census, suggests that
the low-mass PMS stars peak at 
$\alpha_{J2000}=81.3\arcdeg, \delta_{J2000}=+1.5\arcdeg$,
$23.6 \arcmin$ south-east of the star 25 Ori,
with a maximum surface density of $\sim 128 \> stars/deg^2$. However, the
PMS stars with $V_r$ values within $1\sigma$ of the cluster peak velocity 
concentrate slightly north of the density maximum.
We argue that this may indicate that the cluster does indeed extend
further north. \\
We derive a cluster radius of $\sim 7$ pc. 
In a simple-minded kinematic evolution scenario, this value is consistent with
the 25 Ori group representing the evolved counterpart of a $\sim 4$ Myr
old aggregate like the $\sigma$ Ori cluster.

Our results for the 25 Ori cluster in Orion OB1a provide the first large, 
well-defined sample of $\sim 10$ Myr-old stars suitable for studies of 
stellar and disk evolution.

\acknowledgments

C. Brice\~no acknowledges support from grant S1-2001001144 of
FONACIT, Venezuela. This work has been supported by
NSF grant AST-9987367 and NASA grant NAG5-10545.
We are grateful to Susan Tokarz at CfA, who is in charge of
the reduction and processing of FAST, Hectospec and Hectochelle spectra.
We thank the invaluable assistance of the observers and night
assistants, in particular Freddy Moreno, Orlando Contreras, Ubaldo Sanchez
and Gregory Rojas, at the Venezuela Schmidt telescope that made
possible obtaining the data over these past years.
We also acknowledge the support from the
CIDA technical staff, and in particular of Gerardo S\'anchez.

{\it Facilities:} \facility{MMT ()}

\clearpage

\begin{deluxetable}{ccclcccc}
\tabletypesize{\small}
\tablecaption{Hectochelle Observations\label{hcobs}}
\tablehead{
\colhead{Field} & \colhead{$\rm \alpha_{(J2000)}$} & \colhead{$\rm \delta_{(J2000)}$} & \colhead{Location} & \colhead{No. TTS} & \colhead{No. TTS} & \colhead{Airmass} & \colhead{$\rm T_{exp}$} \\
\colhead{} & \colhead{} & \colhead{} & \colhead{} & \colhead{} & \colhead{with fibers} & \colhead{} & \colhead{(sec)}
}
 \startdata
  1   & $81.3033\arcdeg $ & $+1.662\arcdeg $ & 25 Ori &     124  &   81  &  1.46 & $3 \times$ 1200 \\
  2   & $83.2675\arcdeg $ & $-1.614\arcdeg $ & Ori OB1b &   119  &   68  &  1.26 & $3 \times 900$ \\
\enddata
\end{deluxetable}

\clearpage

\begin{deluxetable}{rcccclrrrrrrcc}
 \tabletypesize{\tiny}
 \tablewidth{0pt}
 \tablecaption{Low-mass pre-main sequence members in our 25 Orionis and Ori OB1b Hectochelle fields\label{ttsdata}}
 \tablehead{
\colhead{CVSO}& \colhead{RA(2000)} & \colhead{DEC(2000)}& \colhead{V}&
\colhead{$\rm V-I_C$}& \colhead{SpT}\tablenotemark{1}& \colhead{$\rm W(H\alpha)$\tablenotemark{1}}& 
\colhead{$\rm W(Li~ I)$\tablenotemark{1}} &
\colhead{$\rm W(H\alpha)_{HC}$\tablenotemark{2}}& \colhead{$\rm V_r$\tablenotemark{2}}& \colhead{$\rm \sigma(V_r)$\tablenotemark{2}}& 
\colhead{R} & \colhead{Type} & \colhead{Location} \\
\colhead{}& \colhead{hh:mm:ss} & \colhead{$\rm ^\circ \> \arcmin \> \arcsec$}& \colhead{mag}&
\colhead{mag}& \colhead{} & \colhead{(\AA)} & \colhead{(\AA)}& \colhead{(km/s)}&
\colhead{km/s}& \colhead{km/s} & \colhead{}& \colhead{}& \colhead{}
 }
\startdata
   19 & 05:23:30.93    & +01:45:49.3    & 15.87 &  2.44 &  M3   &   -5.4 &   0.4 &   -4.70 &   17.6 &   1.5 &   5.9 & W   & 25Ori  \\ 
  198 & 05:23:46.57    & +01:45:15.8    & 13.97 &  1.27 &  K4   &   -0.1 &   0.5 &   -0.08 &   19.1 &   0.5 &  17.8 & W   & 25Ori  \\ 
  199 & 05:23:55.36    & +01:34:18.7    & 16.92 &  2.40 &  M3   &   -4.4 &   0.3 &   -5.21 &   20.1 &   1.4 &   5.8 & W   & 25Ori  \\ 
  200 & 05:24:02.30    & +01:53:21.5    & 16.54 &  2.36 &  \nodata  &   -5.1 &  \nodata &   -5.12 &   19.1 &   1.2 &   5.7 & W   & 25Ori  \\ 
   24 & 05:24:04.26    & +01:48:30.5    & 16.34 &  2.22 &  M2   &   -6.1 &   0.3 &   -5.67 &   19.3 &   0.9 &   9.1 & W   & 25Ori  \\ 
  201 & 05:24:07.56    & +01:38:41.7    & 16.28 &  2.17 &  M3   &   -3.2 &   0.2 &   -6.07 &   19.9 &   1.7 &   3.4 & W   & 25Ori  \\ 
   25 & 05:24:10.36    & +01:44:07.6    & 15.42 &  1.88 &  M0   &   -2.6 &   0.3 &   -1.42 &   19.6 &   0.7 &  13.4 & W   & 25Ori  \\ 
  202 & 05:24:18.14    & +01:39:22.9    & 17.41 &  2.61 &  M3   &   -4.7 &   0.5 &   -4.44 &   18.8 &   1.8 &   3.1 & W   & 25Ori  \\ 
  203 & 05:24:20.27    & +01:40:54.8    & 16.66 &  2.37 &  M3   &   -5.6 &   0.4 &   -3.37 &   19.0 &   1.4 &   5.9 & W   & 25Ori  \\ 
  204 & 05:24:25.45    & +01:48:14.3    & 17.45 &  2.72 &  M4   &   -6.6 &   0.6 &   -6.37 &   18.0 &   1.7 &   3.2 & W   & 25Ori  \\ 
  205 & 05:24:26.64    & +01:18:13.9    & 14.77 &  1.99 &  M2   &   -2.1 &   0.3 &   -2.08 &   25.1 &   0.7 &  12.0 & W   & 25Ori  \\ 
  206 & 05:24:41.04    & +01:54:38.6    & 14.93 &  1.54 &  K5   &  -53.5 &   0.5 &  -43.42 &   18.8 &   0.9 &  11.4 & C   & 25Ori  \\ 
  207 & 05:24:52.27    & +01:38:43.9    & 13.63 &  1.16 &  K4   &   -0.8 &   0.5 &   -0.61 &   24.2 &   0.8 &  15.0 & W   & 25Ori  \\ 
  208 & 05:24:54.18    & +01:34:02.1    & 18.08 &  3.22 &  M5   &  -10.0 &   0.6 &   -3.51 &   19.9 &   1.8 &   3.1 & W   & 25Ori  \\ 
  209 & 05:25:01.92    & +01:34:56.4    & 16.87 &  2.56 &  M3   &   -5.8 &   0.5 &   -4.74 &   18.5 &   1.2 &   5.6 & W   & 25Ori  \\ 
   29 & 05:25:02.07    & +01:37:21.2    & 16.19 &  2.49 &  M3   &   -7.6 &   0.5 &   -7.13 &   22.1 &   1.1 &   7.1 & W   & 25Ori  \\ 
  210 & 05:25:03.61    & +01:44:12.1    & 16.92 &  2.46 &  M3   &   -5.9 &   0.3 &   -4.15 &   19.6 &   1.6 &   5.2 & W   & 25Ori  \\ 
  211 & 05:25:08.76    & +01:45:54.3    & 13.98 &  1.24 &  K5   &   -0.1 &   0.5 &   -0.13 &   19.4 &   0.5 &  17.6 & W   & 25Ori  \\ 
  212 & 05:25:13.55    & +01:16:20.6    & 17.55 &  2.72 &  M3   &   -6.9 &   0.5 &   -3.64 &   21.4 &   2.1 &   3.5 & W   & 25Ori  \\ 
  213 & 05:25:16.72    & +01:16:15.9    & 16.08 &  2.53 &  M3   &   -4.4 &   0.2 &   -4.07 &   -0.1 &   2.1 &   5.2 & W   & 25Ori  \\ 
  214 & 05:25:17.04    & +01:48:22.4    & 13.42 &  1.04 &  K2   &   -0.2 &   0.5 &   -0.06 &   19.1 &   0.5 &  17.6 & W   & 25Ori  \\ 
  215 & 05:25:27.70    & +01:41:46.9    & 17.54 &  2.63 &  M3.5 &   -8.0 &   0.2 &   -4.28 &   17.4 &   2.0 &   4.0 & W   & 25Ori  \\ 
  216 & 05:25:34.25    & +01:34:05.7    & 18.45 &  2.95 &  M4.5 &   -7.1 &   0.3 &   -7.02 &   17.5 &   1.6 &   4.1 & W   & 25Ori  \\ 
  217 & 05:25:34.39    & +01:52:19.8    & 15.29 &  1.92 &  M1   &   -2.1 &   0.3 &   -1.70 &   19.2 &   0.7 &  13.2 & W   & 25Ori  \\ 
  218 & 05:25:35.18    & +01:43:40.6    & 17.29 &  2.62 &  M4   &   -7.7 &   0.3 &   -4.31 &   19.0 &   1.7 &   4.2 & W   & 25Ori  \\ 
  219 & 05:25:35.65    & +01:28:23.2    & 15.05 &  1.82 &  M0   &   -1.3 &   0.5 &   -0.13 &   21.0 &   0.6 &  13.4 & W   & 25Ori  \\ 
  220 & 05:25:38.89    & +02:01:29.9    & 14.01 &  1.19 &  K4   &   -0.7 &   0.4 &   -0.40 &   19.0 &   0.7 &  15.8 & W   & 25Ori  \\ 
  221 & 05:25:39.57    & +01:42:53.7    & 15.10 &  1.94 &  M0   &   -1.8 &   0.5 &   -1.53 &   20.6 &   0.7 &  12.4 & W   & 25Ori  \\ 
  222 & 05:25:39.72    & +01:25:46.9    & 16.03 &  2.49 &  M3   &   -9.9 &   0.3 &   -7.47 &   15.5 &   1.9 &   6.1 & W   & 25Ori  \\ 
  223 & 05:25:39.94    & +01:40:21.1    & 16.50 &  2.41 &  M3   &   -5.3 &   0.4 &   -4.35 &   21.2 &   1.1 &   6.7 & W   & 25Ori  \\ 
   35 & 05:25:45.91    & +01:45:50.2    & 14.72 &  1.91 &  K7   &  -10.3 &   0.4 &  -15.72 &   18.8 &   1.0 &  11.0 & C   & 25Ori  \\ 
  224 & 05:25:46.74    & +01:43:30.4    & 17.21 &  2.60 &  M3   &  -20.3 &   0.3 &  -13.67 &   18.4 &   1.7 &   3.1 & W   & 25Ori  \\ 
   36 & 05:25:50.37    & +01:49:37.3    & 16.00 &  2.22 &  M3   &   -8.9 &   0.5 &   -3.92 &   17.9 &   0.8 &  10.7 & W   & 25Ori  \\ 
  225 & 05:25:51.82    & +01:47:36.1    & 16.17 &  2.58 &  M4   &   -4.9 &   0.2 &   -3.45 &   18.5 &   1.3 &   5.8 & W   & 25Ori  \\ 
  226 & 05:25:52.61    & +01:34:44.2    & 17.75 &  2.73 &  M4   &   -4.9 &   0.3 &   -3.98 &   20.4 &   1.3 &   4.5 & W   & 25
Ori  \\ 
  227 & 05:25:59.05    & +01:23:34.7    & 16.30 &  2.60 &  M4   &   -4.8 &   0.2 &   -4.99 &   20.2 &   1.2 &   6.0 & W   & 25Ori  \\ 
  228 & 05:26:06.36    & +01:45:46.7    & 16.30 &  2.26 &  M2   &   -3.4 &   0.2 &   -3.84 &   20.6 &   0.9 &   8.8 & W   & 25Ori  \\ 
   38 & 05:26:06.41    & +01:37:11.8    & 16.18 &  2.26 &  M2   &   -7.5 &   0.5 &   -4.04 &   20.8 &   1.1 &   9.1 & W   & 25Ori  \\ 
  229 & 05:26:07.84    & +01:23:08.3    & 15.95 &  2.43 &  M3   &   -4.3 &   0.2 &   -3.71 &   24.1 &   1.6 &   5.5 & W   & 25Ori  \\ 
  230 & 05:26:08.02    & +01:41:15.2    & 15.44 &  1.94 &  M2   &   -4.9 &   0.5 &   -2.90 &   19.9 &   1.1 &  11.6 & W   & 25Ori  \\ 
  231 & 05:26:10.10    & +01:18:44.3    & 16.42 &  2.69 &  M4   &   -9.0 &   0.2 &   -7.28 &   21.4 &   1.7 &   3.9 & W   & 25Ori  \\ 
  232 & 05:26:11.18    & +01:54:52.3    & 14.70 &  1.94 &  M2   &   -3.7 &   0.3 &   -3.22 &   19.1 &   1.2 &  10.5 & W   & 25Ori  \\ 
  233 & 05:26:21.98    & +01:47:57.6    & 16.30 &  2.46 &  M3   &   -5.0 &   0.3 &   -4.46 &   18.3 &   1.2 &   6.8 & W   & 25Ori  \\ 
  234 & 05:26:38.04    & +01:21:45.5    & 15.35 &  2.08 &  M2   &   -4.0 &   0.4 &   -3.77 &   24.0 &   0.8 &  10.4 & W   & 25Ori  \\ 
  235 & 05:26:42.65    & +01:39:47.3    & 16.65 &  2.33 &  M2.5 &   -5.7 &   0.3 &   -7.69 &   20.3 &   1.3 &   5.5 & W   & 25Ori  \\ 
   42 & 05:26:50.43    & +01:55:03.6    & 15.99 &  2.16 &  M2   &   -3.8 &   0.3 &   -2.15 &   20.7 &   0.9 &   8.7 & W   & 25Ori  \\ 
  236 & 05:27:00.52    & +01:49:27.2    & 16.51 &  2.36 &  M3.5 &   -7.7 &   0.3 &   -5.45 &   29.6 &   1.1 &   8.2 & W   & 25Ori  \\ 
   89 & 05:31:19.99    & -01:46:53.3    & 15.47 &  1.57 &  K7   &   -1.6 &   0.4 &   -1.18 &   30.5 &   0.7 &  13.3 & W   & OB1b   \\ 
   92 & 05:31:31.75    & -01:47:05.6    & 16.90 &  2.21 &  M2   &   -3.6 &   0.2 &   -5.53 &   23.6 &   1.9 &   3.7 & W   & OB1b   \\ 
   97 & 05:31:42.80    & -01:39:48.3    & 16.79 &  2.41 &  M2   &   -5.4 &   0.2 &   -6.46 &   30.4 &   1.8 &   4.8 & W   & OB1b   \\ 
  237 & 05:31:48.29    & -01:47:59.2    & 16.36 &  2.36 &  M3   &   -4.3 &   0.4 &   -3.78 &   30.8 &   0.8 &  10.5 & W   & OB1b   \\ 
   99 & 05:31:50.79    & -01:55:17.5    & 17.37 &  2.44 &  M2   &   -5.5 &   0.2 &   -5.06 &   30.9 &   1.1 &   8.2 & W   & OB1b   \\ 
  238 & 05:32:00.42    & -01:40:11.0    & 15.85 &  1.88 &  \nodata  &  -47.5 &  \nodata &  -47.46 &   32.4 &   0.8 &  13.1 & C   & OB1b   \\ 
  239 & 05:32:03.47    & -01:56:32.0    & 14.91 &  1.56 &  K5   &   -1.2 &   0.3 &   -1.31 &    4.6 &   0.6 &  16.3 & W   & OB1b   \\ 
  240 & 05:32:05.41    & -01:46:31.8    & 17.42 &  2.47 &  \nodata  &   -5.7 &   0.2 &   -4.16 &   29.9 &   1.7 &   4.0 & W   & OB1b   \\ 
  106 & 05:32:21.51    & -01:43:45.6    & 16.80 &  2.44 &  M2   &   -7.0 &   0.3 &   -8.39 &   31.0 &   2.1 &   3.4 & W   & OB1b   \\ 
  241 & 05:32:25.69    & -01:33:11.8    & 14.92 &  1.67 &  K6   &   -3.8 &   0.6 &   -3.23 &   29.9 &   0.8 &  13.0 & C   & OB1b   \\ 
  242 & 05:32:28.93    & -01:27:34.6    & 17.85 &  2.06 &  M1   &   -0.2 &   0.2 &   -2.03 &   55.6 &   2.1 &   3.2 & W   & OB1b   \\ 
  243 & 05:32:29.90    & -01:52:40.0    & 15.37 &  1.69 &  K6   &   -0.9 &   0.5 &   -0.61 &   31.5 &   0.6 &  16.7 & W   & OB1b   \\ 
  110 & 05:32:34.35    & -01:50:46.3    & 15.41 &  1.60 &  M0   &   -2.0 &   0.6 &   -1.72 &   31.7 &   0.7 &  12.5 & W   & OB1b   \\ 
  244 & 05:32:59.21    & -02:00:38.1    & 16.00 &  2.47 &  M1   &   -4.8 &   0.5 &   -4.45 &   20.5 &   0.8 &   9.8 & W   & OB1b   \\ 
  245 & 05:33:06.62    & -01:47:03.2    & 17.60 &  2.79 &  M3   &   -6.6 &   0.3 &   -5.01 &   33.0 &   1.7 &   5.1 & W   & OB1b   \\ 
  246 & 05:33:10.69    & -01:31:35.8    & 15.06 &  2.17 &  \nodata  &   -2.5 & \nodata &   -2.54 &   25.5 &   0.9 &  11.6 & W   & OB1b   \\ 
  247 & 05:33:19.87    & -01:56:28.9    & 17.57 &  2.37 &  M2   &   -2.8 &   0.2 &   -2.63 &   32.3 &   1.1 &   7.7 & W   & OB1b   \\ 
  248 & 05:33:22.72    & -01:43:42.2    & 16.04 &  1.91 &  K7   &   -3.3 &   0.6 &   -2.12 &   27.6 &   0.8 &  11.8 & W   & OB1b   \\ 
  118 & 05:33:25.76    & -01:32:18.7    & 14.70 &  1.26 &  K5   &   -1.0 &   0.5 &   -0.72 &   31.6 &   0.6 &  15.9 & W   & OB1b   \\ 
  249 & 05:33:29.30    & -01:30:28.2    & 17.00 &  2.69 &  M3   &   -5.5 &   0.3 &   -6.05 &   32.9 &   1.9 &   4.2 & W   & OB1b   \\ 
  250 & 05:33:36.11    & -01:28:52.6    & 15.00 &  1.69 &  K6   &   -1.3 &   0.5 &   -1.30 &   29.8 &   0.7 &  15.8 & W   & OB1b   \\ 
  251 & 05:33:41.87    & -01:47:40.7    & 16.14 &  1.94 &  M0   &   -3.3 &   0.5 &   -8.35 &   29.7 &   0.8 &  11.5 & W   & OB1b   \\ 
  252 & 05:33:43.63    & -01:13:36.2    & 17.29 &  2.48 &  M3   &   -3.2 &   0.3 &   -3.11 &   27.9 &   1.8 &   4.0 & W   & OB1b   \\ 
  253 & 05:33:46.84    & -01:56:55.1    & 15.01 &  1.49 &  K3   &   -0.5 &   0.4 &   -0.63 &   31.5 &   0.6 &  17.2 & W   & OB1b   \\ 
  122 & 05:33:47.29    & -01:20:27.3    & 15.85 &  1.85 &  M1   &   -1.9 &   0.5 &   -1.66 &   29.8 &   0.8 &  11.6 & W   & OB1b   \\ 
  254 & 05:34:00.70    & -01:50:08.4    & 15.70 &  1.78 &  K6   &   -2.4 &   0.5 &   -2.06 &   31.2 &   0.8 &  12.4 & W   & OB1b   \\ 
  255 & 05:34:06.17    & -01:58:55.9    & 16.75 &  2.39 &  M3   &   -4.4 &   0.2 &   -2.98 &   26.7 &   2.2 &   5.1 & W   & OB1b   \\ 
  256 & 05:34:15.31    & -01:15:19.6    & 16.57 &  2.46 &  M3   &   -3.8 &   0.4 &   -3.82 &   30.2 &   1.7 &   5.6 & W   & OB1b   \\ 
  257 & 05:34:15.46    & -01:34:29.8    & 17.77 &  2.80 &  M4   &   -7.3 &   0.5 &   -7.35 &   27.6 &   2.1 &   3.3 & W   & OB1b   \\ 
  258 & 05:34:17.87    & -01:41:08.0    & 14.57 &  1.50 &  K4   &   -1.6 &   0.5 &   -0.56 &   27.2 &   0.7 &  14.1 & W   & OB1b   \\ 
  134 & 05:34:38.98    & -01:46:56.4    & 15.25 &  1.71 &  M0   &   -4.6 &   0.5 &   -2.84 &   29.0 &   0.6 &  14.9 & W   & OB1b   \\ 
\enddata
\tablenotetext{1}{Measurements from low-resolution spectra (FAST, Hydra, Hectospec).}
\tablenotetext{2}{Measurements from the high-resolution Hectochelle spectra.}
\tablecomments{The classification in column {\it Type} (C=CTTS, W=WTTS) is based on spectral dependent
threshold $\rm W(H\alpha)$ following White \& Basri (2003).
Spectral type errors are 1 subclass.
Stars CVSO-200, 238 and 246 have no low-resolution spectra available yet. They were classified 
as T Tauri stars from the Hectochelle spectra based on the following criteria: having the appropriate 
magnitudes and colors, radial velocity within 1-$\sigma$ of the peak of the velocity distribution
for their location, peculiar H$\alpha$ emission, either a blue-shifted absorption component to
H$\alpha$ or this line very strongly in emission (the later is the case for CVSO-238).\\
Table \ref{ttsdata}  is published in its entirety in the electronic version
of the Astrophysical Journal. A portion is shown here for guidance regarding
its form and contents.}
\end{deluxetable}

\clearpage

\clearpage

\clearpage

\clearpage

\clearpage

\end{document}